\documentclass[conference]{IEEEtran}
\IEEEoverridecommandlockouts
\usepackage{cite}
\usepackage{amsmath,amssymb,amsfonts}
\usepackage{braket}
\usepackage{booktabs}
\usepackage{algorithmic}
\usepackage{algorithm2e}
\usepackage{graphicx}
\usepackage{textcomp}
\usepackage{comment}
\usepackage{xcolor}
\usepackage{amsthm}
\usepackage{soul}
\usepackage{tabto}

\usepackage{tikz}
\usepackage{verbatim}
\newcommand\copyrighttext{%
  \footnotesize \textcopyright 2026 IEEE. Personal use of this material is permitted.
  Permission from IEEE must be obtained for all other uses, in any current or future
  media, including reprinting/republishing this material for advertising or promotional
  purposes, creating new collective works, for resale or redistribution to servers or
  lists, or reuse of any copyrighted component of this work in other works.}
\newcommand\copyrightnotice{%
\begin{tikzpicture}[remember picture,overlay]
\node[anchor=south,yshift=10pt] at (current page.south) 
  {\fbox{\parbox{\dimexpr\textwidth-\fboxsep-\fboxrule\relax}{\copyrighttext}}};
\end{tikzpicture}%
}

\def\BibTeX{{\rm B\kern-.05em{\sc i\kern-.025em b}\kern-.08em
    T\kern-.1667em\lower.7ex\hbox{E}\kern-.125emX}}
\begin{document}

\title{Bit-Vector Abstractions to Formally Verify Quantum Error Detection \& Entanglement}

\author{\IEEEauthorblockN{Arun Govindankutty}
\IEEEauthorblockA{\textit{Electrical \& Computer Engineering,} \textit{North Dakota State University,
Fargo, ND USA}
}}

\maketitle
\copyrightnotice
\begin{abstract}
As the number of qubits increases, quantum circuits become more complex and their state space grows rapidly. This makes functional verification challenging for conventional techniques. Ensuring correctness is especially critical for quantum error correction and entanglement generation. This paper presents a novel application of bit-vector based abstraction methodology for formal verification of quantum circuits where superposition and functional behaviour can be decoupled. The approach is applied to error detection circuits for 2-qubit, 3-qubit, and Shor 9-qubit quantum codes, as well as Bell-state and GHZ-state generation circuits. The error detection circuits and the Bell-state generation circuit are verified in less than a second and 25MB memory. GHZ circuits with up to 8,192 qubits are verified in under three minutes using a maximum of 23.2,GB of memory. The results demonstrate the versatility, scalability, and effectiveness of the proposed approach.

\end{abstract}


%\vspace{-0.1in}
\section{Introduction}
%\vspace{-0.05in}
\label{sec:intro}
Quantum computing has the potential to outperform classical computing in several application domains. These include communication, machine learning, material science, medicine, and optimization, and collectively motivate the pursuit of quantum advantage~\cite{IET_futureQC, qc_arun, qvsm_iet_breast_cancer, wireless_networks_optimization, tqe_portfolio}. Quantum algorithms exploit uniquely quantum phenomena, such as superposition and entanglement, to solve problems that are intractable for classical computers.
Entanglement of qubits plays a central role in the realization of multi-qubit quantum systems and the effective execution of quantum algorithms. Also, current quantum hardware is highly susceptible to errors~\cite{Shor_fault_tolerant_QC, Shor_error_code, Brooks2019, Preskill2018}. This vulnerability limits the reliability of quantum computations. Therefore, verifying circuits that generate entanglement and provide error resilience is essential for the practical advancement of quantum computing.

Verifying quantum circuits is challenging due to several inherent factors. Qubit superposition, state collapse during measurement, and the exponential growth of the Hilbert space significantly increase verification complexity. As a result, traditional verification techniques become inefficient for quantum circuits. Formal methods can address these challenges by using abstractions to enable reliable, robust, and efficient verification of quantum systems~\cite{quantum_hoare_logic, quantum_formal_survey, arun_review}.

This work proposes a formal methodology for the functional verification of quantum circuits. The method is applied to error detection circuits of various quantum error correction codes, Bell-state generation, and Greenberger-Horne-Zeilinger (GHZ) state preparation. The key contributions of this work are summarized as follows:
\begin{itemize}
    \item A bit-vector abstraction that reduces quantum error detection, Bell-state generation, and GHZ-state preparation circuits from the Hilbert space representation to a bit-vector domain.
    \item Functional abstractions of quantum gates tailored for quantum error detection, Bell-state generation, and GHZ-state preparation circuits.
    \item Formal specification and verification of correctness properties for quantum error detection, Bell-state generation, and GHZ-state preparation circuits.
\end{itemize}

The proposed approach provides a scalable and rigorous framework for the formal verification of quantum circuits.

%%\vspace{-0.05in}
\section{Theoretical Background}
%\vspace{-0.05in}
\label{sec:bckgnd}
This section provides a brief background on qubits, quantum gates and circuits, quantum errors, and syndrome extraction. For a more comprehensive understanding of quantum computing fundamentals, readers may refer to~\cite{qc_textbook, qc_book_mit}.
In classical computing, information is processed using \textit{bits}, which take values of 0 or 1. In quantum computing, however, information is represented using \textit{qubits} (short for quantum bits). Mathematically, a qubit is expressed as a vector in Dirac notation: $|\psi\rangle = \alpha|0\rangle + \beta|1\rangle$, where 
$|\psi\rangle$ represents the quantum state of the qubit. $|0\rangle$ and $|1\rangle$ denote the computational basis vectors. The coefficients $\alpha,\beta \in \mathbb{C}$, $|\alpha|^2$ and $|\beta|^2$ are the probability amplitudes, constrained by the normalization condition $\alpha^2 + \beta^2 = 1$. Since the computational basis vectors are orthonormal, a qubit state is a linear superposition of these basis states. The individual computational basis vectors are defined as follows:
%\vspace{-0.2in}
%\begin{center}
\begin{equation}
|0\rangle =
\begin{bmatrix}
1\\
0
\end{bmatrix}, 
\text{and} \ \ |1\rangle =
\begin{bmatrix}
0\\
1
\end{bmatrix}
\label{qubit_eqn}
\end{equation}
%\end{center} 

In quantum computing, computation is carried out by evolving the quantum states of qubits over time. This evolution is governed by unitary operators, commonly referred to as quantum gates. A quantum circuit consists of an ordered sequence of such gates. It specifies the step-by-step operations used to solve a given problem. Accordingly, a quantum circuit can be viewed as a quantum program or a quantum algorithm, as it encodes the logical structure of the computation~\cite{quantum_ckt}.

A multi-qubit system with $n$ qubits is represented by a state vector in a $2^n$-dimensional Hilbert space. The quantum state is expressed as
\begin{equation}
    \ket{\Psi} = \sum_{x \in \{0,1\}^n} \alpha_x \ket{x}.
\end{equation}
The exponential growth of this state space is a key source of quantum computational power. It also makes verification of quantum circuits complex.

Measurement projects the quantum state onto a basis state. For a measurement in the computational basis, the probability of observing an outcome $x$ is given by
\begin{equation}
    \Pr(x) = |\alpha_x|^2.
\end{equation}
This inherent probabilistic behaviour makes reasoning about correctness challenging. It motivates the use of formal methods that can reason over distributions of measurement outcomes.

\subsection{Quantum Entanglement}
Entanglement is a uniquely quantum correlation between subsystems. 
A multi-qubit state is entangled if it cannot be expressed as a tensor product of individual qubit states. For example, the Bell-state
\begin{equation}
    \ket{\Phi^+} = \frac{1}{\sqrt{2}} (\ket{00} + \ket{11})
\end{equation}
exhibits maximal entanglement. Figure~\ref{fig:ckt_fig} (b) elucidates the quantum circuit for this Bell state generation.

The GHZ state is an entangled quantum state involving three or more qubits~\cite{greenberger2007goingbellstheorem}. The generalized GHZ state consists of $m>2$ subsystems. For a system of $m$ qubits (qubits imply two-dimensional system), the GHZ state is given by
\begin{equation}
    \ket{\textbf{GHZ}} = \frac{1}{\sqrt{2}} \left( \ket{0}^{\otimes m} + \ket{1}^{\otimes m} \right),
\end{equation}
where $\ket{0}^{\otimes m}$ and $\ket{1}^{\otimes m}$ denote the basis states $\ket{00\ldots0}$ and $\ket{11\ldots1}$, respectively. Figure~\ref{fig:ckt_fig}(c) shows the GHZ-state generation for 4-qubit system.

Entanglement is a key resource for many quantum algorithms. It violates classical assumptions of locality and independence. As a result, classical verification techniques are inadequate, which motivates the development of quantum-specific formal reasoning frameworks.

%\vspace{-0.05in}
\subsection{Quantum Errors and Error Correcting Codes}
%\vspace{-0.05in}
Errors in an algorithm or a program occur when the output deviates from the expected result for a given input. In quantum circuits, these errors originate from complex physical interactions governing the system~\cite{qec_beginners, qec_book}. They can arise due to coherent quantum errors caused by incorrect gate applications, environmental decoherence, qubit loss, state leakage, measurement inaccuracies, and initialization errors. To detect and mitigate these errors, qubit encoding techniques have been developed, enabling reliable quantum error detection and correction. Among the most widely studied error detection codes are the 2-qubit codes~\cite{gottesman_qec_intro}, 3-qubit codes~\cite{qec_intro}, and the Shor 9-qubit code~\cite{shor_9_qubit}. 
In all qubit encoding schemes, the logical qubit $|\psi\rangle_L$ is represented using multiple physical qubits. For instance, in the 2-qubit code, the logical qubit is expressed as $|\psi\rangle_L = |0\rangle_L + |1\rangle_L$, where $|0\rangle_L \mapsto |00\rangle$, and $|1\rangle_L \mapsto |11\rangle$. Similarly, in the 3-qubit code, the logical basis states are defined as $|0\rangle_L \mapsto |000\rangle$, and $|1\rangle_L \mapsto |111\rangle$. The 9-qubit code extends this concept by encoding each logical qubit using three repetitions of the 3-qubit code, with entanglement introduced through controlled-NOT (C-NOT) gates to enhance error detection and correction capabilities. In general, 2-qubit, 3-qubit, and 9-qubit codes can correct a \textit{single qubit flip error}. Additionally, the 9-qubit code has the capability to correct a \textit{single phase flip error} occurring in any one of the 3-qubit logical blocks.

\begin{figure}[!ht]
\centering
\includegraphics[width=.48\textwidth]{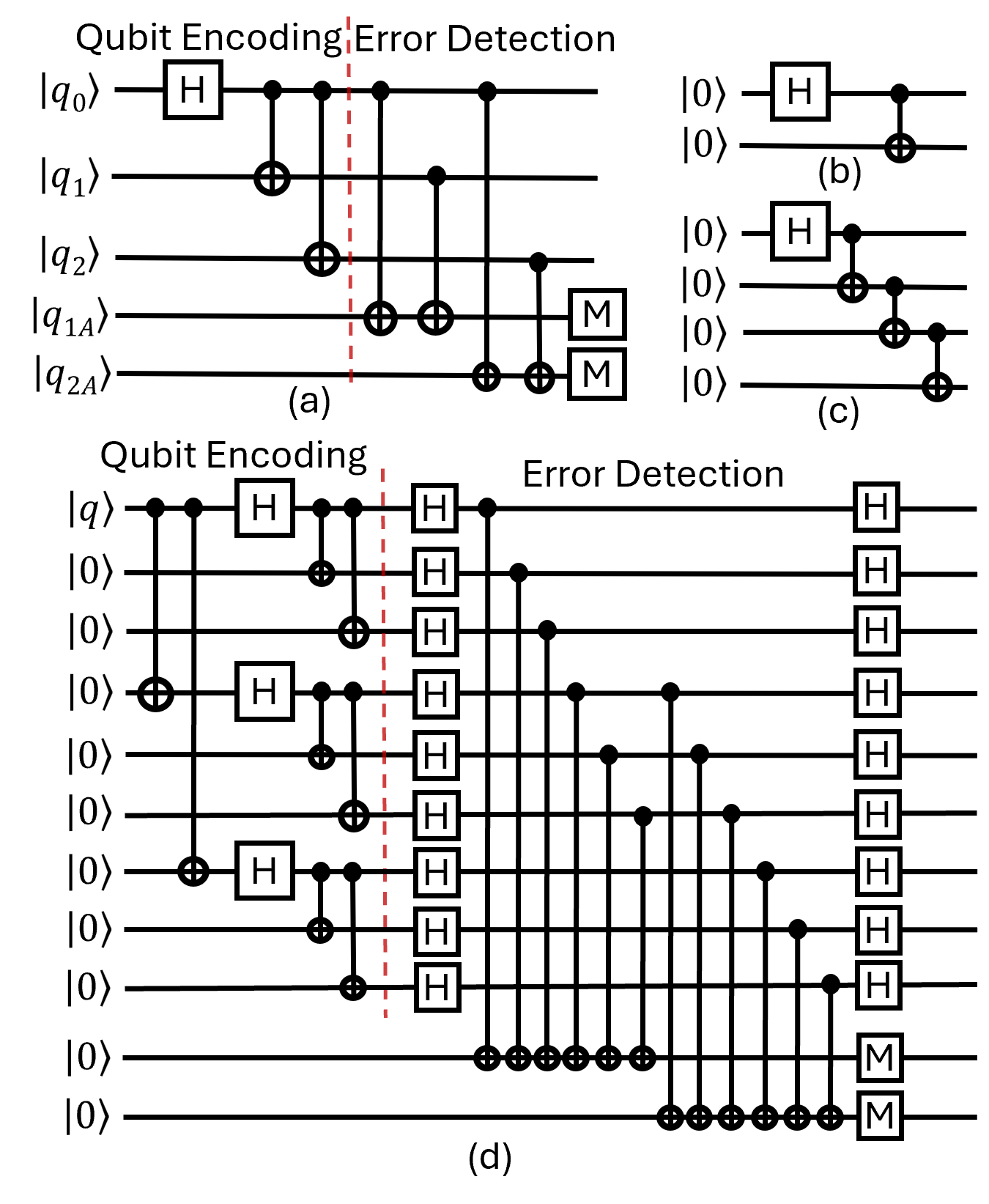}
\vspace{-0.1in}
\caption{Figure shows quantum circuits for the following: (a) 3-qubit code with error detection. (b) Bell state ($\frac{1}{\sqrt{2}}(\ket{0}+\ket{1})$). (c) GHZ-state with 4-qubits ($\frac{1}{\sqrt{2}}(\ket{0000}+\ket{1111})$). (d) 9-qubit Shor code with error detection. H indicates Hadamard gate and M indicates measurement operation on qubits.}
%\vspace{-0.3cm}
\label{fig:ckt_fig}
\end{figure} 

The error detection circuits are designed to detect errors in a quantum system while preserving the original qubit states. Figure~\ref{fig:ckt_fig}(a) shows the 3-qubit error detection code along with the error detection circuit. %Figure~\ref{3q_correct} (a) illustrates the circuit implementation for the 3-qubit code, including the error detection circuit with ancilla qubits, implemented using IBM Qiskit~\cite{Qiskit}. Figure~\ref{3q_correct} (b) presents the probability distribution of the correct logical qubit states along with the measured values of the ancilla qubits. 
In this circuit, qubits $q_0, q_1, q_2$ form the logical qubit $|\psi\rangle_L$, while qubits $q_{1_A}$ and $q_{2_A}$ serve as ancilla qubits, $|\psi\rangle_A$. For simplicity, we combine both the ancilla and the logical qubits into a single representation. The combined representation is as follows:
$|\psi\rangle_A |\psi\rangle_L = |00\rangle |000\rangle = |00000\rangle$. 

For the 3-qubit code, when no errors occur, the ancilla qubits should hold the state `00'. However, in the presence of a qubit-flip error, the ancilla qubits can take values `01', `10', or `11', as summarized in Table~\ref{tab:syn_code}.
\vspace{-0.1in}
\begin{table}[!h]
    \centering
    \caption{Error code Table for 3-qubit code}
    \begin{tabular}{c c c}
    \toprule
    Ancilla Value ($q_{2_A}q_{1_A}$) & Qubit State & Inference \\ 
    \midrule
    00 & $|000\rangle+|111\rangle$ & No Error \\
    11 & $|001\rangle+|110\rangle$ & $1^{st}$ qubit flip \\
    01 & $|010\rangle+|101\rangle$ & $2^{nd}$ qubit flip  \\
    10 & $|100\rangle+|011\rangle$ & $3^{rd}$ qubit flip  \\ 
    \bottomrule
    \end{tabular}
    \label{tab:syn_code}
\end{table}
An error detection circuit consists of Hadamard (H) gates and C-NOT gates. The matrix representation of the gates are given below.
\begin{equation}
\text{H} = \frac{1}{\sqrt{2}}
\begin{bmatrix}
1 & 1\\
1 & e^{\pi i}
\end{bmatrix} = \frac{1}{\sqrt{2}}
\begin{bmatrix}
1 & 1\\
1 & -1
\end{bmatrix}
\end{equation}
\begin{equation}
\text{C-NOT} =
\begin{bmatrix}
1 & 0 & 0 & 0 \\
0 & 1 & 0 & 0 \\
0 & 0 & 0 & 1 \\
0 & 0 & 1 & 0
\end{bmatrix}
\end{equation}

Figure~\ref{fig:ckt_fig} (d) shows the Shor-9-qubit code along with the quantum circuit for detecting phase flip errors. 
The ancilla qubit values `00, 11, 01 and 10' correspond to no-error, phase flip in the first, second and third logic blocks respectively.  
By projecting the error information onto the ancilla qubits, the circuit ensures that the quantum state of the data qubits remains intact, enabling reliable error detection and correction.

%\vspace{-0.05in}
\subsection{Formal Verification}
%\vspace{-0.05in}
Formal verification is a rigorous validation framework based on mathematical reasoning and proofs. It systematically analyses the entire design space in contrast to traditional testing which analyses specific test scenarios. This enables the detection of corner-case bugs that may be missed by conventional methods. As a result, formal verification provides strong guarantees of correctness.
Formal methods are widely used in software verification, hardware verification, and VLSI design, where reliability and accuracy are critical~\cite{formal_default, formal_silicon, formal_vlsi}. Recent research has focused on extending these techniques to quantum circuit verification~\cite{qc_formal_springer, quantum_formal_survey}. 
%\vspace{-0.05in}
In this work, we propose a formal method to verify quantum circuits for error detection and entanglement generation.

\section{Existing Approaches}
\label{sec:relWorks}

This section provides a concise review of existing formal methods for quantum algorithm verification and highlights the unique contributions of our proposed approach. While current verification techniques focus on validating various quantum algorithms, they do not specifically address the critical challenge of error detection circuit and entanglement verification. To the best of our knowledge, this work is the first to tackle this essential requirement, making it highly relevant in the NISQ (Noisy Intermediate-Scale Quantum) era. By filling this gap, our approach ensures more reliable and scalable verification of quantum error detection mechanisms.

Amy~\cite{qc_path_sum} proposes a verification methodology based on dyadic arithmetic and rewrite rules using the Haskell theorem prover. In this approach as qubit count increases, integer overflow issues arise, leading to verification failures. The largest quantum circuit verified using this method contains 96 qubits, whereas our proposed approach is capable of verifying circuits with up to 8,192 qubits demonstrating significantly greater scalability.

Similarly, Seiter \emph{et al.}~\cite{decision_diag} present a property-checking method utilizing multiple-valued decision diagrams. While they successfully verify Grover's search algorithm, their approach is limited to circuits with a maximum of 10 qubits, highlighting significant scalability constraints.

Burgholzer and Wille~\cite{adv_eq_chk_qc}, as well as Yamashita and Markov~\cite{yamashita2010}, propose equivalence checking as a method for quantum circuit verification. Their approach relies on the availability of an already verified reference circuit to perform the verification process. Also, when a circuit cannot be reduced to binary space, their method resorts to a hybrid approach that incorporates Hilbert space representations, which significantly limits scalability. As a result, their technique can verify circuits with a maximum of 128 qubits, posing challenges for larger-scale quantum systems. Our approach does not require a verified reference circuit and reduces the entire problem to bit-vector space providing scalability upto 8,192 qubits.

Govindankutty \emph{et al.}~\cite{qft_arun, iet_qenc_arun, iqft_arun} introduced symbolic abstractions to reduce quantum circuits to a bit-vector representation. Their approach was applied to the verification of Quantum Fourier Transform, its inverse circuits, and quantum data encoding circuits only. They do not address error detection or entanglement generation.
In this work, we build on these abstractions and extend their scope. The proposed framework supports the verification of a broader class of quantum circuits, including quantum error detection codes and quantum entanglement generation circuits. Through this work, we demonstrate the versatility of symbolic abstractions and formal methods for scalable quantum circuit verification.

%\vspace{-0.05in}

\vspace{-0.1in}
\section{Verification Framework}
%\vspace{-0.05in}
\label{sec:frmWrk}
This section presents the abstraction definitions and properties introduced in this work. As discussed in Section~\ref{sec:bckgnd}, a qubit is expressed as a linear combination of computational basis states. Therefore, if a circuit is verified to be correct for all basis states, its correctness extends to the entire state space spanned by those states. This observation forms the foundation of the proposed abstraction framework.

%\vspace{-0.05in}
\subsection{Abstractions}
\label{sec:abstractions}
%\noindent
\textbf{Definition~1 (Abstract Qubit).}
 \textit{An abstract qubit $Q$ is a bit-vector tuple $(s, q)$, where $s$ is a 3-bit vector that encodes the superposition status of the qubit, and $q$ is a 2-bit vector that encodes the measurement status and the computational basis vector value.}

In this abstraction, $s \in \{b'000, b'001, \ldots, b'111\}$ captures the superposition characteristics of the qubit. The value of $q$ represents the computational basis state and whether qubit is measured. When $q \in \{b'00, b'01\}$, the qubit is unmeasured. When $q \in \{b'10, b'11\}$, the qubit has been measured. The most significant bit of $q$ indicates the measurement status, and the least significant bit stores the measured basis value. The prefix $b'$ denotes a bit-vector (binary) encoding.

\textbf{Definition~2 (Abstract $H$ Gate).}  
\textit{The abstract Hadamard gate $H$ is a state transition function on an abstract qubit $Q(s, q)$:  
\begin{equation}
H(Q) = 
\begin{cases} 
(s \ +_{2} \ b'001, q) & \text{if } \forall s \in \{b'000,\dots,b'110\} \\ 
(b'110, q) & \text{if }  s = b'111
\end{cases}
\end{equation}
}
In this abstraction framework, even values of $s$ indicate that the qubit is not in superposition (i.e., zero or an even number of $H$-gates applied). Odd values of $s$ indicate the qubit is in superposition due to a single or odd number of $H$-gate applications. The $s$ value thus tracks the number of $H$-gates applied by modulo-2 ($+_2$) addition of $b'001$. Value $b'000$ means no $H$-gates, $b'001$ means one $H$-gate, and so on. For more than seven applications, $s$ toggles between $b'110$ and $b'111$ as the length of $s$ is 3 bits.  
In this work, entanglement circuit uses one $H$-gate and the error detection circuit uses at most two $H$-gates. Thus 3 bits are sufficient for $s$ ($s > b'010$ is an error). The $H$-gate does not modify the $q$ component of the abstract qubit.

\textbf{Definition~3 (Abstract $X$ Gate).}  
\textit{The abstract $X$-gate is a transformation on an abstract qubit $Q(s, q)$, negating the input basis bit without affecting superposition bits. 
\begin{equation}
X(Q(s,q)) = Q(s,\neg q)
\end{equation}
}
Here, $Q$ represents the abstract qubit, and $\neg$ represents logical negation.

\textbf{Definition~4 (Abstract C-NOT Gate).}  
\textit{The abstract C-NOT gate is a transformation on a control-target qubit pair $(Q_c, Q_t)$:  
\begin{equation}
\mathcal{C}(Q_c, Q_t) = 
\begin{cases} 
(Q_c, Q_t) & \text{if } q_{c_0} = 0 \\ 
(Q_c, Q(s_t, \neg q_{t_0})) & \text{if } q_{c_0} = 1 
\end{cases}
\end{equation}
}

Here, $Q_c$ and $Q_t$ denote the abstract control and target qubits, and $\neg$ represents logical negation. The operation leaves the control qubit unchanged. The target qubit basis bit $q_{t_0}$ is inverted only if the control bit $q_{c_0}$ is 1; otherwise, it remains the same. The measurement indicators ($q_{c_1}, q_{t_1}$) and the superposition indicators ($s_c, s_t$) are unchanged. This abstraction captures the conditional behaviour of the C-NOT gate while preserving the qubit's abstract state.

\textbf{Definition~5 (Singular Function).}  
\textit{If $m = \{m_0, m_1, \dots, m_{n-1}\}$ be a bit vector of length $n$, where $m_i \in \{b'0, b'1\} \forall i \in [0, n)$, then, Singular Function $\mathcal{S}$ is defined as:  
\begin{equation}
\mathcal{S}(m) = 
\begin{cases} 
b'1 & \text{if } \sum_{i=0}^{n-1} m_i = b'1 \\ 
b'0 & \text{otherwise} 
\end{cases}
\end{equation}
}

The function $\mathcal{S}(m)$ returns $b'1$ if exactly one bit in the vector is set to $b'1$. Otherwise, it outputs $b'0$. This function is essential for verifying the correctness of error detection circuits.

\subsection{Properties}
We define the correctness properties next. Separate properties are defined for bit-flip error, phase-flip error detection, and entanglement generation verification. Abstraction aligned properties are defined next.

\textbf{Property~1 (Bit-flip Correctness).}  
\textit{$\forall \ (s^{\text{in}}_{A_i} = b'000) \wedge (s^{\text{in}}_{L_i} = b'000)$, the following conditions must hold:  
\begin{enumerate}
    \item $(s^{\text{out}}_{A_i} = b'000) \wedge (s^{\text{out}}_{L_i} = b'000)$,
    \item $(q^{\text{in}}_{A_i} = b'00) \wedge (q^{\text{out-1}}_{A_i} = b'0D)$,
    \item $(q^{\text{out}}_{A_0} = b'11 \Leftrightarrow q_{L_{00}} \neq q_{L_{10}}) \wedge (q^{\text{out}}_{A_1} = b'11 \Leftrightarrow q_{L_{00}} \neq q_{L_{20}})$.
\end{enumerate}
}

This property enforces three key conditions:  
First, both ancilla and logical qubits must preserve their superposition states throughout the error detection process. No unintended change to the superposition should occur.  
Second, ancilla qubits must be properly initialized, and should not be measured until the final output. The symbol $D$ indicates dont-care that is the basis state may be either 0 or 1.  
Third, ancilla qubits are measured only at the output and activated only in the presence of a single bit-flip error. This ensures that each error detection circuit functions correctly, guaranteeing the circuit’s reliability against `bit-flip' errors.

\textbf{Property~2 (Phase-flip Correctness).}  
\textit{$\forall \ (s^{\text{in}}_{A_i} = b'000) \wedge (s^{\text{in}}_{L_i} = b'000)$, the following conditions must hold:  
\begin{enumerate}
    \item $(s^{\text{out}}_{A_i} = b'000) \wedge (s^1_{L_i} = b'001) \wedge (s^{\text{out}}_{L_i} = b'010)$,
    \item $(q^{\text{in}}_{A_i} = b'00) \wedge (q^{\text{out-1}}_{A_i} = b'0D)$,
    \item $\mathcal{S}(q_{L_0}, q_{L_1}, \dots , q_{L_8}) \implies$  
    \[
    \langle q^{\text{out}}_{A_1} q^{\text{out}}_{A_0} \rangle = 
    \begin{cases} 
    b'1011 & \text{if } \mathcal{S}(q_{L_{0}}, q_{L_{1}}, q_{L_{2}}) = 1, \\
    b'1111 & \text{if } \mathcal{S}(q_{L_{3}}, q_{L_{4}}, q_{L_{5}}) = 1, \\ 
    b'1110 & \text{if } \mathcal{S}(q_{L_{6}}, q_{L_{7}}, q_{L_{8}}) = 1.
    \end{cases}
    \]
\end{enumerate}
}

This property enforces three key conditions:  
First, superposition correctness must be maintained. Ancilla qubits preserve their superposition states, while logical qubits contain exactly one Hadamard ($H$) gate at the beginning and one at the end. This ensures that the $s$ component of each logical qubit remains at $b'001$ during extraction.  
Second, ancilla qubits must be properly initialized and remain unmeasured until the final output stage. The symbol $D$ indicates dont care, that is, the qubit’s basis state may be 0 or 1.  
Third, ancilla qubits reflect single-qubit phase-flip errors in the logical qubits. If exactly one of $q_{L_0}, q_{L_1}, \dots, q_{L_8}$ is flipped, the ancilla qubits $q_{A_1}$ and $q_{A_0}$ indicate the affected logical block. For example, $b'1010$ indicates no error, while errors in the first, second, or third 3-qubit blocks correspond to $b'1011$, $b'1111$, and $b'1110$, respectively. This mapping enables precise phase-flip error detection within the Shor 9-qubit code.

\textbf{Property~(Entanglement Correctness).}  
\textit{For a quantum system $Q=\{Q_1, Q_2, \dots, Q_m\}$ to be entangled, the following conditions must hold:
\begin{enumerate}
    \item $(s^{\text{in}}_1 = b'000) \wedge (s^{1}_1 = b'001) \wedge (s^{\text{out}}_1 = b'001)$,
    \item $(s^{\text{in}}_2 = s^{\text{in}}_3 = \dots = s^{\text{in}}_m = b'000) \wedge (s^{\text{out}}_2 = s^{\text{out}}_3 = \dots = s^{\text{out}}_m = b'000)$,
    \item $\forall \ i,j \in \{1,2,\dots,m\}, \ (\text{if}, \ i=1 \wedge \ q_i^{in} = b'00) \vee  (\text{if}\ i\neq1, \ \wedge \ q_1^{in} = b'01 \wedge q_i^{in}=b'00) \ \text{then}, (q^{\text{out}}_i = q^{\text{out}}_j$).
\end{enumerate}
}

This property enforces three requirements. First, the initial qubit $Q_1$ must enter superposition through a single Hadamard ($H$) gate applied as the first operation. Its superposition state is $b'000$ at input and remains $b'001$ after the first stage and throughout the circuit.  
Second, the remaining qubits $Q_2,\dots,Q_m$ must never enter superposition. No $H$-gate is applied to these qubits, and their superposition state remains $s=b'000$.  
Third, all qubits must share the same output basis state when the first qubit is initialized to either $\ket{0}$ or $\ket{1}$ and all remaining qubits are initialized to $\ket{0}$ (i.e., $q^{\text{in}}_1 \in \{b'00, b'01\}$ and $q^{\text{in}}_i = b'00$ for all $i \neq 1$). The output must satisfy
$q^{\text{out}}_1 = q^{\text{out}}_2 = \dots = q^{\text{out}}_m$,
which represents the states $\ket{00\ldots0}$ or $\ket{11\ldots1}$. Note that this condition accurately captures the requirement of equal superposition of the basis states for the first qubit by application of $H$-gate required for entanglement generation as well.

\section{Verification Methodology}
\label{sec:methodology}
The quantum circuit under test is abstracted from Hilbert space to bit-vector space using the abstractions (Section~\ref{sec:abstractions}). The abstract circuit is encoded in SMT (Satisfiability Modulo Theory) and verified for functional correctness. Errors are injected into the abstract circuit to evaluate the proposed methodology.  
The correctness of the approach is formally established through the lemmas presented below. These lemmas show that the method can detect errors in quantum circuits for both error detection and entanglement generation.

\textbf{Lemma~1.}  
If any ancilla qubit is initialized incorrectly or subjected to a $H$-gate, then Property~1 (bit-flip correctness) and Property~2 (phase-flip correctness) are violated.

\textbf{Lemma~2.}  
If a Hadamard gate is missing from or incorrectly applied to any logical qubit, then Property~1 (bit-flip correctness) and Property~2 (phase-flip correctness) are violated.

\textbf{Lemma~3.}  
For bit-flip error detection, any misconfiguration of CNOT gates in the circuit, including incorrect control or target qubits, missing, or extra CNOT gates, result in a violation of Property~1.

\textbf{Lemma~4.}  
For phase-flip error detection, any misconfiguration of CNOT gates in the circuit, including incorrect control or target qubits, missing, or extra CNOT gates, result in a violation of Property~2.

\textbf{Lemma~5.}  
Any misconfiguration in Bell-state or GHZ-state generation circuits violates Property~3. This includes missing or extra $H$ gates, missing or extra C-NOT gates, and incorrect control or target qubits in C-NOT gates that result in a non-entangled state.

\textbf{Lemma~6.}  
Any combination of the above lemmas violates at least one of the properties~1--3, flagging every error in functional verification.

\textbf{Theorem~1.} 
\textit{An error in the error detection circuit results in the violation of one or both of Properties 1 and 2 during verification of the abstracted circuit, indicating incorrect functionality.}

\textbf{Theorem~2.} 
\textit{An error in the entanglement generation circuit results in the violation of property 3 during the verification of the abstracted circuit, indicating incorrect functionality.}

Proofs for the above lemmas and theorems are straightforward and omitted for brevity.

\section{Experimental Results}
\label{sec:results}
Experiments for generating the benchmarks, including the abstractions, circuits, and properties are encoded in the SMT-LIB language. The properties were verified using the Z3 Theorem Prover (version 4.14.1)~\cite{z3_cite}. All experiments were conducted on an Intel\textregistered Core\texttrademark Ultra 9 285K CPU running at 3.2 GHz, 192 GB of RAM, and RHEL 9 (64-bit), running the RHEL-9.5 64-bit operating system. 
%\vspace{-0.1in}
\begin{table}[!h]
    \centering
    \caption{Verification Results for Error Detection}
    %\vspace{-0.05in}
    \begin{tabular}{c c c}
    \toprule
    Quantum Code & Time(s) & Memory(MB) \\ 
    \midrule
    2-qubit bit-flip & 0.01 & 17.7 \\
    3-qubit bit-flip & 0.01 & 17.7 \\
    9-qubit phase-flip & 0.01 & 25.7\\ 
    Bell state generation & 0.01 & 17.7 \\
    \bottomrule
    \end{tabular}
    \label{tab:verif_err}
\end{table}
Table~\ref{tab:verif_err} summarizes the verification results for the error detection and the Bell-state generation circuits. The \emph{Quantum Code} column identifies the circuit type. Columns \emph{Time} and \emph{Memory} report the execution time the peak memory usage in seconds and megabytes, respectively. All circuits were verified in 0.01s. The memory usage for each verification run remained below 25MB.
%\vspace{-0.1in}
\begin{table}[!h]
\caption{Verification Results for Entanglement Generation}
\begin{center}
\label{tab:verif_entanglement}
\resizebox{.48\textwidth}{!}{
\begin{tabular}{c c c c c} \toprule
  \textbf{Qubit Count}  & \multicolumn{2}{c} {\textbf{No-Error}}  & \multicolumn{2}{c} {\textbf{Error}} \\  
  &\textbf{Time}(s)  & \textbf{Memory}(MB) &\textbf{Time}(s)  & \textbf{Memory}(MB)\\  
  \toprule
  4 & 0.01 & 17.4 & 0.01 & 17.7\\
  8 & 0.01 & 17.4 & 0.01 & 17.7\\
  16 & 0.01 & 17.5 & 0.01 & 17.8\\
  32 & 0.01 & 17.7 & 0.01 & 18.0\\
  64 & 0.01 & 18.6 & 0.01 & 18.7\\
  128 & 0.02 & 23.0 & .02 & 23.0\\
  256 & 0.05 & 39.9 & 0.05 & 39.9\\
  512 & 0.24 & 115 & 0.25 & 115\\
  1024 & 1.25 & 393 & 1.28 & 393\\
  2048 & 6.71 & 1,624 & 6.68 & 1,624\\
  4096 & 32.7 & 6,173 & 32.0 & 6,173\\
  8192 & 138 & 23,677 & 139 & 23,677\\
  \bottomrule

\end{tabular}
}
\end{center} 
%\vspace{-0.1in}
\end{table}

GHZ-state (entanglement) circuits were verified with qubit counts ranging from 4 to 8,192 to demonstrate scalability of the proposed methodology. Table~\ref{tab:verif_entanglement} reports the corresponding verification results. The \emph{Qubit Count} column indicates the number of qubits in the circuit. The \emph{Time} and \emph{Memory} columns report the execution time in seconds and the peak memory usage in megabytes for both the correct and erroneous circuits. The reported error corresponds to a control-qubit error at qubit-2. The execution time and memory usage are comparable for correct and erroneous circuits. All other error scenarios were also verified and exhibited similar resource usage. These results are omitted from reporting for brevity.
The proposed approach is currently applicable to class of quantum circuits in which superposition can be decoupled from functional behaviour.
The results demonstrate the efficiency, scalability, and robustness of the proposed approach.

\section{Conclusion and Future Directions}
\label{sec:conclusion}

This paper presents a bit-vector–based abstraction for formally verifying quantum circuits in which superposition can be decoupled from functional behaviour. The framework reduces quantum circuit verification from the Hilbert space representation to a bit-vector domain using abstract qubit and gate models. Its versatility is demonstrated by verifying error detection circuits for multiple quantum error-correcting codes, as well as Bell-state and GHZ-state generation circuits. Scalability and efficiency are shown by low verification time, modest memory usage, and support for large qubit counts.

Future work will focus on extending the proposed abstraction framework to support more general classes of quantum circuits, including those involving stabilizer codes, as well as broader categories of quantum algorithms. Such extensions are expected to improve the reliability and scalability of verification techniques for fault-tolerant quantum computing. Furthermore, the results demonstrate the potential of SMT solvers for large scale quantum circuit verification and encourage continued research into automated formal methods for quantum systems.

%\clearpage
%\input{DefProp_Backup}
%\input{Experiments_Backup}
\section*{Acknowledgment}
The author thanks Dr. Kushal Ponugoti for the engaging discussions while working on the quantum error detection section, and Dr. Benjamin D. Braaten for providing the computational resources used in the experiments.

\bibliographystyle{IEEEtran}
\bibliography{ref}

\end{document}